\documentclass[prd,aps,nofootinbib,showpacs,onecolumn, 11 pt]{revtex4}
\usepackage{graphicx}
\usepackage{epsfig}

\begin{document}

\title{$\Lambda_b \to p, \,\, \Lambda$ transition form factors from QCD light-cone sum rules}

\author{Yu-Ming Wang$^{a}$, Yue-Long Shen$^{b}$ and  Cai-Dian L\"u$^{a}$}

\affiliation{
 \it $^a$ Institute of High Energy Physics and Theoretical Physics Center for Science Facilities,
  P.O. Box 918(4) Beijing 100049,  China\\
 \it $^b$  College of Information Science and Engineering,
Ocean University of China, Qingdao, Shandong 266100, P.R. China }

\date{\today}

\begin{abstract}

Light-cone sum rules for the $\Lambda_b \to p, \,\, \Lambda$
transition form factors are derived from the correlation functions
expanded by the twist of the distribution amplitudes of $\Lambda_b$
baryon. In terms of the $\Lambda_b$ three-quark distribution
amplitudes models constrained by the QCD theory, we calculate the
form factors at small momentum transfers and compare the results
with that estimated in the conventional light-cone sum rules (LCSR)
and perturbative QCD approaches. Our results indicate that the two
different version of sum rules can lead to the consistent numbers of
form factors responsible for $\Lambda_b \to p$ transition. The
$\Lambda_b \to \Lambda$ transition form factors from
 LCSR with the asymptotic $\Lambda$ baryon distribution amplitudes
are found to be almost one order larger than that obtained in the
$\Lambda_b$-baryon LCSR, implying that the pre-asymptotic
corrections to the baryonic distribution amplitudes are of great
importance. Moreover, SU(3) symmetry breaking effect between the
form factors $f_1^{\Lambda_b \to p}$ and $f_1^{\Lambda_b \to
\Lambda}$ are computed as $28^{+14}_{-8} \%$ in the framework of
$\Lambda_b$-baryon LCSR.

\end{abstract}


\maketitle


\section{Introduction}

The priority to investigate $b$ quark decays can be attributed to
their sensitivity of the flavor structure of nature, which serves as
a touch-stone in the ongoing effort to explore the standard model
(SM) describing the interactions between elementary particles. Weak
decays of heavy baryons containing a $b$ quark may provide important
clues on flavor-changing currents beyond the SM in a complementary
fashion to $B$ meson decays. $\Lambda_b \to p, \, \Lambda$
transition form factors are the essential hadronic objects in the
exclusive semileptonic $\Lambda_b \to p l \bar{\nu}_l$,  $\Lambda_b
\to \Lambda l \bar{l}$ and radiative $\Lambda_b \to \Lambda \gamma$
decays. Such form factors can  also be employed to describe the
nonleptonic charmless  $\Lambda_b$ decays in terms of factorization
approach. There is no doubt that  reliable estimation of the
transition form factors in QCD is indispensable to an accurate
determination of the flavor-changing couplings of the quarks (which
is known as Cabibbo-Kobayashi-Maskawa (CKM) matrix elements) and to
a deep understanding of the underlying structure and dynamics of
hadrons and currents.

There has been continuous interest concentrated on the theoretical
analysis of form factors from the underlying field theory, the main
challenge of which is to deal with the nonperturbative effect in the
hadron as a bound state  properly. Several theoretical tools going
beyond the realm of perturbation theory  have been developed in this
aspect, such as lattice QCD (LQCD), perturbative QCD (PQCD)
approach, QCD sum rules (QCDSR) and light-cone sum rules (LCSR)
approaches. LQCD is only applicable to the computations of
heavy-to-light transition form factors with large momentum transfer.
 Phenomenologically, PQCD approach has been applied
to the investigations of various baryonic transitions including
proton Dirac form factor \cite{Li:1992ce,Kundu:1998gv}, semileptonic
charmless decay $\Lambda_b \to p l\bar{\nu}$ \cite{Shih:1998pb},
semileptonic charming decay $\Lambda_b \to \Lambda_c l\bar{\nu}$
\cite{Shih:1999yh,Guo:2005qa} and  radiative decay $\Lambda_b \to
\Lambda \gamma$ \cite{He:2006ud}. Three-point QCDSR approach based
on the short-distance operator product expansion (OPE) and double
dispersion relations have also been employed to calculate the weak
transition form factors, such as $\Lambda_b \to p l\bar{\nu}$
\cite{Huang:1998rq} and $\Lambda_b \to \Lambda \gamma, \Lambda l
\bar{l}$ \cite{Huang:1998ek}, where the nonperturbative
contributions are embedded in the vacuum condensates of quarks and
gluons.

Alternative sum rule approach to hadronic form factors is to perform
the OPE of a dedicated correlation function near the light-cone and
the light-cone distribution amplitudes are employed to describe the
long-distance dynamics in the correlator. As a marriage of the
standard QCDSR technique and the theory of hard exclusive process,
the LCSR procedure involves a partial resummation of local operators
and cure the problem of QCDSR applied to large momentum transfer. An
attractive advantage of the LCSR is that it offers an systematic way
to take into account both hard scattering and soft (end-point)
contributions to the transition form factors almost model
independently \cite{Colangelo:2000dp}. Utilizing the distribution
amplitudes of proton and $\Lambda$ baryon, the form factors
responsible for  $\Lambda_b \to p l\bar{\nu}$ \cite{Huang:2004vf}
and $\Lambda_b \to \Lambda \gamma, \Lambda l \bar{l}$
\cite{Wang:2008sm} transitions have been investigated in the LCSR
approach. The main uncertainties in the standard LCSR approach
originate from the less known nonperturbative parameters involved in
the distribution amplitudes of light hadrons, apart from the
systematic uncertainty brought by the quark-hadron duality
assumption in the heavy hadron channel.

Another version of LCSR approach ($B$ meson LCSR) proposed  in Ref.
\cite{Khodjamirian:2005ea,Khodjamirian:2006st} starts with the
$B$-to-vacuum correlation function, where the light hadron is
interpolated by an appropriate current and the $B$ meson is put on
the mass shell. The on-shell $B$ meson can be well described in the
heavy quark effective theory (HQET) and the correlation function is
found to be light-cone dominance. In this context, the distribution
amplitudes of $B$ meson are treated as universal nonperturbative
inputs, which embody the long-distance dynamics of all the $B \to L$
($L$ being a light meson) transition form factors. Different from
the LCSR with light-meson distribution amplitudes (light-meson
LCSR), systematic uncertainty  in $B$ meson LCSR is from the
quark-hadron duality in the channels of light mesons. Along this
line, LCSR with distribution amplitudes of heavy hardons have been
employed to compute the form factors of $B \to D^{(\ast)}$
\cite{Faller:2008tr}, $B \to a_1(1260)$ \cite{Wang:2008bw}, and
$\Lambda_b \to \Lambda_c$ \cite{Wang:2009ym} very recently. In the
present work, we would like to follow the same prescription to
analyze the $\Lambda_b \to p, \,\, \Lambda$ transition form factors,
which provides an independent way  to test  predictions of the
standard  LCSR with light-hadron distribution amplitudes.

The fundamental nonperturbative functions in the $\Lambda_b$ baryon
LCSR are the distribution amplitudes of $\Lambda_b$, which describes
the hadronic structure in rare parton configurations with a fixed
number of Fock components at small transverse separation in the
infinite momentum frame. The complete classification of three-quark
distribution amplitudes of $\Lambda_b$ baryon in the heavy quark
limit has been carried out in Ref. \cite{Ball:2008fw}, where the
evolution equation for the leading twist distribution amplitude is
also derived.  It has been shown that the evolution equation for the
``leading-twist" distribution amplitude contains one piece related
to the Lange-Neubert kernel \cite{Lange:2003ff}, which generates a
radiative tail when either of the two momenta $\omega_{1,2}$ is
large, and another piece related to the ERBL kernel
\cite{Lepage:1979zb}, which redistributes the momenta within the
spectator di-quark system.

The layout of this paper is as follows: In section II,  we collect
the distribution amplitudes of $\Lambda_b$ baryon to the leading
Fock state. Parameterizations of the various hadronic matrix element
$\langle L(P) | \bar{q} \gamma_{\mu} b| \Lambda_b (P+q)\rangle $
($L$ denotes a light baryon) with $\Gamma_i$ being all the possible
Lorentz structures are presented in section III, where the
applicability of OPE on the light cone for the correlation function
is also briefly reviewed. The sum rules for the $\Lambda_b \to L$
transition form factors up to the twist-4 are then derived. It is
shown that the relation of form factor in the heavy quark limit are
well respected in the $\Lambda_b$ baryon LCSR. Numerical analysis of
LCSR for the transition form factors at large recoil region are
displayed in section IV, where detailed comparisons of the results
with that obtained in the conventional LCSR, three-point QCDSR and
PQCD approaches are also discussed. In particular, the estimation of
SU(3) symmetry breaking effect between the form factors
$f_1^{\Lambda_b \to p}$ and $f_1^{\Lambda_b \to \Lambda}$ are
included  here. The last section is devoted to the conclusion.

\section{Distribution amplitudes of $\Lambda_b$ baryon}

The complete set of three-quark distribution amplitudes of
$\Lambda_b$ baryon in the heavy quark limit can be constructed
as\cite{Ball:2008fw}
\begin{eqnarray}
\epsilon^{abc} \langle 0 | u^a(t_1 n) C\gamma_5 \not n d^b (t_2 n)
h_v^c(0) | \Lambda(v) \rangle &=& f^{(2)}_{\Lambda} \Psi_2(t_1, t_2)
\Lambda(v) \, , \nonumber  \\
\epsilon^{abc} \langle 0 | u^a(t_1 n) C\gamma_5  d^b (t_2 n)
h_v^c(0) | \Lambda(v) \rangle &=& f^{(1)}_{\Lambda} \Psi_3^{s}(t_1,
t_2) \Lambda(v) \, ,\nonumber  \\
\epsilon^{abc} \langle 0 | u^a(t_1 n) C\gamma_5 i \sigma_{\bar n n}
d^b (t_2 n) h_v^c(0) | \Lambda(v) \rangle &=& 2 f^{(1)}_{\Lambda}
\Psi_3^{\sigma}(t_1, t_2) \Lambda(v) \, , \nonumber  \\
\epsilon^{abc} \langle 0 | u^a(t_1 n) C\gamma_5 \not \bar{n} d^b
(t_2 n) h_v^c(0) | \Lambda(v) \rangle &=& f^{(2)}_{\Lambda}
\Psi_4(t_1, t_2) \Lambda(v), \label{Lambdab distribution amplitudes}
\end{eqnarray}
where the subscript $2$, $3$, $4$ refers to the twist of the diquark
operator, $t_i$ are arbitrary real numbers, describing the locations
of the valence quarks inside the $\Lambda_b$ baryon on the
light-cone. The light-like unit vectors $n_{\mu}$ and
$\bar{n}_{\mu}$ satisfy $n^2=\bar{n}^2=0$, $n \cdot \bar{n}=2$,
therefore the four-velocity of the $\Lambda_b$ baryon
$v_{\mu}=(n_{\mu} +\bar{n}_{\mu}) /2$. The gauge links between the
fields in the above have been suppressed for brevity.

It is straightforward to rewrite the Eq. (\ref{Lambdab distribution
amplitudes}) in the following form
\begin{eqnarray}
&& \epsilon^{abc} \langle 0 | u^a_{\alpha}(t_1 n) d^b_{\beta} (t_2
n) {h_v}^c_{\gamma}(0) | \Lambda(v) \rangle  \nonumber \\ \nonumber
&& = {1 \over 8} f_{\Lambda}^{(2)} \Psi_2(t_1, t_2) (\not \bar{n}
\gamma_5 C)_{\alpha \beta} \Lambda_{\gamma}(v) + {1 \over 4}
f_{\Lambda}^{(1)} \Psi_3^s (t_1, t_2) ( \gamma_5 C)_{\alpha \beta}
\Lambda_{\gamma}(v)
\nonumber \\
&&-{1 \over 8} f_{\Lambda}^{(1)} \Psi_3^{\sigma} (t_1, t_2) ( i
\sigma_{\bar n n} \gamma_5 C)_{\alpha \beta} \Lambda_{\gamma}(v) +{1
\over 8} f_{\Lambda}^{(2)} \Psi_4(t_1, t_2) (\not n \gamma_5
C)_{\alpha \beta} \Lambda_{\gamma}(v). \label{Lambdab nonlocal
matrix elements}
\end{eqnarray}
Each distribution amplitude $\Psi_i(t_1, t_2)$ can be expressed by a
Fourier integral
\begin{eqnarray}
\Psi_i(t_1, t_2) =  \int_0^{+\infty}  d\omega_1 \int_0^1 d\omega_2
e^{- i \omega_1 t_1 u - i t_2 \omega_2} \psi_i(\omega_1, \omega_2) =
\int_0^{+\infty} \omega d\omega \int_0^1 du e^{- i \omega(t_1 u +t_2
\bar{u})} \tilde{\psi}_i(\omega,u),
\end{eqnarray}
where $\tilde{\psi}(\omega,u)=\psi(u\omega, \bar{u} \omega)$,
$\bar{u}=1-u$, $\omega_i$ ($i=1, \, 2$) are the energies of the $u-$
and $d-$ quarks. $\omega=\omega_1+\omega_2$ is the total energy
carried by light quarks and the dimensionless parameter $u$
describes the momentum fraction carried by the $u$ quark in the
diquak system. The normalization of $\tilde{\psi}_i$ are
\begin{eqnarray}
\int_0^{+\infty} \omega d\omega \int_0^1 \tilde{\psi}_2(\omega,u)=
\int_0^{+\infty} \omega d\omega \int_0^1
\tilde{\psi}_3^{s}(\omega,u)= \int_0^{+\infty} \omega d\omega
\int_0^1 \tilde{\psi}_4(\omega,u)=1.
\end{eqnarray}

The explicit forms of the distribution amplitudes for the
$\Lambda_b$  baryon  are proposed as
\begin{eqnarray}
\tilde{\psi}_2(\omega,u) &=& \omega^2 u (1-u) \bigg [ {1 \over
\epsilon_0^4 } e^{- \omega /\epsilon_0}  + a_2 C_2^{3 /2}(2u-1) {1
\over \epsilon_1^4} e^{- \omega /\epsilon_1}  \bigg], \nonumber \\
\tilde{\psi}_3^s(\omega,u) &=& { \omega \over 2 \epsilon_3^3} e^{-
\omega /\epsilon_3}, \nonumber \\
\tilde{\psi}_3^{\sigma}(\omega,u) &=& { \omega \over 2 \epsilon_3^3}
(2
u-1) e^{-\omega /\epsilon_3}, \nonumber \\
\tilde{\psi}_4(\omega,u) &=& 5 \mathcal{N}^{-1}
\int_{\omega/2}^{s_0^{\Lambda_b}} ds \,\, e^{- s / \tau} (s-
\omega/2)^3 ,
\end{eqnarray}
where $\tau$ is taken to be in the interval $0.4 < \tau < 0.8 {\rm
GeV}$, $s_0^{\Lambda_b}=1.2 {\rm GeV}$ is the continuum threshold
for the $\Lambda_b$ channel in the heavy-quark effective theory
(HQET) and the coefficient
\begin{eqnarray}
\mathcal{N}= \int_0^{s_0^{\Lambda_b}} s^5 e^{-s/\tau} .
\end{eqnarray}

\section{Light-cone sum rules for the transition form factors}

\subsection{Parameterizations of transition form factors}

Generally, the hadronic matrix elements responsible for $\Lambda_b$
decays to a light baryon $L$ can be parameterized in terms of a
series of form factors
\begin{eqnarray}
\label{parameterization of vector current 1} \langle L(P) | \bar{q}
\gamma_{\mu} b| \Lambda_b (P+q)\rangle = \bar{L}(P) (f_1
\gamma_{\mu} + f_2 i \sigma_{\mu \nu} q ^{\nu}+ f_3
q_{\mu} ) \Lambda_b (P+q) ,  \\
\langle L(P) | \bar{q} \gamma_{\mu} \gamma_5 b| \Lambda_b
(P+q)\rangle = \bar{L}(P) (F_1 \gamma_{\mu} + F_2 i \sigma_{\mu \nu}
q ^{\nu}+ F_3 q_{\mu} ) \gamma_5 \Lambda_b (P+q) ,
\label{parameterization of vector current 2}
\\
\label{parameterization of tensor current 1}
\langle L(P) | \bar{q}
i \sigma_{\mu \nu } q^{\nu} b| \Lambda_b (P+q)\rangle = \bar{L}(P)
(g_1 \gamma_{\mu} + g_2 i \sigma_{\mu \nu}
q ^{\nu}+ g_3 q_{\mu} )  \Lambda_b (P+q) , \\
\langle L(P) | \bar{q} i \sigma_{\mu \nu } q^{\nu} \gamma_5 b|
\Lambda_b (P+q)\rangle = \bar{L}(P) (G_1 \gamma_{\mu} + G_2 i
\sigma_{\mu \nu} q ^{\nu}+ G_3 q_{\mu} )  \gamma_5 \Lambda_b (P+q),
\label{parameterization of tensor current 2}
\end{eqnarray}
where all the form factors  $f_i$ ,  $F_i$ , $g_i$ and $G_i$ are
functions of  the momentum transfer $q^2$.

For the completeness, we also present the parameterizations of
matrix elements involving the scalar $\bar{q} b$ and pseudo-scalar
$\bar{q} \gamma_5 b$ currents, which can be obtained from  Eqs.
(\ref{parameterization of tensor current 1},\ref{parameterization of
tensor current 2}) by contracting both sides to the four-momentum
$q^{\mu}$
\begin{eqnarray}
\langle L(P)|\bar{q} b|\Lambda_b(P+q)\rangle &=& {1 \over m_b+m_q}
\overline{L}(P) [g_1(m_{\Lambda_b}-m_{L})+g_3 q^2]\Lambda_b(P+q),
\,\, \label{scalar
matrix element}\\
\langle L(P)|\bar{q}\gamma_{5} b|\Lambda_b(P+q)\rangle &=& {1 \over
m_b-m_q} \overline{L}(P) [G_1(m_{\Lambda_b}+m_{L})-G_3
q^2]\gamma_5\Lambda_b(P+q). \,\, \label{pseudo-scalar matrix
element}
\end{eqnarray}

In the heavy quark limit, the form factors $f_i$,  $F_i$ , $g_i$ and
$G_i$ can be expressed by two independent functions $\xi_1$ and
$\xi_2$
\begin{eqnarray}
\langle L(P)|\bar{q} \Gamma b|\Lambda_b(P+q)\rangle = \bar{L}(P)
[\xi_1(q^2) + \not v \xi_2(q^2)] \Gamma \Lambda_B (P+q) \label{form
factors in HQET}
\end{eqnarray}
with $\Gamma$ being an arbitrary Lorentz structure. Comparing Eqs.
(\ref{parameterization of vector current 1})-(\ref{parameterization
of tensor current 2}) and Eq. (\ref{form factors in HQET}), one can
easily get
\begin{eqnarray}
&& f_1 =F_1  = g_2 =G_2 = \xi_1 + { m_L \over m_{\Lambda_b} } \xi_2,
\nonumber  \\
&& g_1=G_1= { q^2 \over m_{\Lambda_b} } \xi_2 , \nonumber  \\
&& f_2=F_2 =f_3 =F_3 = { \xi_2 \over m_{\Lambda_b} } , \nonumber  \\
&& g_3 ={ m_L- m_{\Lambda_b}  \over m_{\Lambda_b} } \xi_2 ,
\nonumber  \\
&& G_3 ={ m_L+ m_{\Lambda_b}  \over m_{\Lambda_b} } \xi_2 ,
\label{relations of form factors in HQET}
\end{eqnarray}
where $m_L$ denotes the mass of light baryon. It is known that Eq.
(\ref{form factors in HQET}) is successful at the small-recoil
region (with large $q^2$) in the heavy quark limit.

\subsection{Correlation functions}

Following Ref. \cite{Khodjamirian:2006st}, the correlation function
of two quark currents relevant to the $\Lambda_b \to L$ transition
is taken between the vacuum and the on-shell $\Lambda_b$ baryon
state
\begin{eqnarray}
z_{\mu} T^{\mu}(P,q) = i  z_{\mu} \int d^4 x e^{i p \cdot x} \langle
0 |T{j_1(x), {j_2}^{\mu} (0)} | \Lambda_b (P+q)\rangle \,\, ,
\label{correlation function}
\end{eqnarray}
where $j_1(x)$ is the interpolating current for a light baryon and
$j_2(0)$ denotes the weak transition current. The introduction of
$z_{\nu}$, satisfying $z^2=0$ and $z \cdot q =0$,  is to remove the
contributions $\sim z^{\nu}$ that give subdominant contributions on
the light-cone. The manifest forms of  currents $j_i \, \, (i=1,2)$
for $\Lambda_b \to p, \Lambda$ transitions and their form factors
are grouped in Table \ref{choices of currents}. The correlation
function can be systematically expanded in terms of the heavy quark
mass in HQET. The momentum of $\Lambda_b$ baryon can be redefined as
$P+q=m_{b} v +k$, where $k$ is the residual momentum and the
relativistic normalization of the state is $| \Lambda_b (P+q)
\rangle = |\Lambda_b (v) \rangle $, up to $1/m_b$ corrections.
Moreover, it is also convenient to rescale the $b-$ quark field by
introducing an effective field $ h_v(x)= b(x) e^{i m_b v x} +
O(1/m_b)$. In the first approximation,
$m_{\Lambda_b}=m_b+\bar{\Lambda}$ implying that $k_0 \sim
\bar{\Lambda}$. Accordingly, the four-momentum transfer $q$ is
redefined by separating the ``static" part of  it: $q= m_b v
+\tilde{q}$, hence we have $p +\tilde{q} =k$. Now, it is straight to
translate the correlation function (\ref{correlation function}) to
HQET
\begin{eqnarray}
z_{\mu} T^{\mu}(P,q) =  z_{\mu} \tilde{T}^{\mu}(P,\tilde{q})
+O(1/m_b)
\end{eqnarray}
where the effective correlation  function
\begin{eqnarray}
z_{\mu} \tilde{T}^{\mu}(P,\tilde{q})= i z_{\mu} \int d^4 x e^{i p
\cdot x} \langle 0 |T{j_1(x), {\tilde{j}_2}^{\mu} (0)} | \Lambda_b
(v)\rangle \label{correlator LC}
\end{eqnarray}
does not depend on the $b$ quark mass. ${\tilde{j}_2}^{\mu} (0)$ can
be obtained from ${j_2}^{\mu} (0)$ by replacing $b$ quark field with
the effective field $h_v$. It can be found that the correlation
function (\ref{correlator LC}) is dominated by the light-cone region
$x^2 \leq 1/P^2$, if both the four-momentum are spacelike $P^2 <0,
\,\, \bar{q}^2$ and sufficiently large
\begin{eqnarray}
|P^2|, \,\, |\tilde{q}^2| \gg \Lambda_{QCD}^2, \bar{\Lambda},
\end{eqnarray}
apart form the requirement that the ratio $\xi = {2 p \cdot k /
P^2}$ should be at least of $O(1)$. In the language of effective
theory, the initial external momentum $q$ can be written as
\begin{eqnarray}
q^2 \approx m_b^2 + 2 m_b \tilde{q}_0 \sim m_b^2 -m_b P^2 \xi
/\bar{\Lambda} .
\end{eqnarray}
As a  result, the operator product expansion (OPE) on the light-cone
works well  in the kinematical region
\begin{eqnarray}
0\leq q^2 < m_b^2- m_b P^2 / \bar{\Lambda} . \label{region for OPE}
\end{eqnarray}

 \begin{table}
 \caption{The two currents involved in the correlation function (\ref{correlation function}) and the corresponding heavy-to-light form factors.}
 \label{choices of currents}
 \begin{center}
\begin{tabular}{cccc}
  \hline
 \hline
  Transition & $j_1$ & $j_2^{\mu}$ & form factors \\
  \hline
  $\Lambda_b \to p$ & $\epsilon^{ijk}[u^i(x) C \not z u^j (x)  ]  \,\, \gamma_5 \not z d^k (x) $ & $\bar{u}(0)\gamma^{\mu} (1- \gamma_5) b(0)$ & $f_i^{\Lambda_b \to p}$, $F_i^{\Lambda_b \to p}$ \\
  $\Lambda_b \to \Lambda$  & $\epsilon^{ijk}[u^i(x) C \gamma_5 \not z d^j (x)  ]  \not z s^k (x) $ & $\bar{s}(0)\gamma_{\mu }(1- \gamma_5) b(0)$ & $f_i^{\Lambda_b \to \Lambda}$, $F_i^{\Lambda_b \to \Lambda}$ \\
  $\Lambda_b \to \Lambda$  & $\epsilon^{ijk}[u^i(x) C \gamma_5 \not z d^j (x)  ]  \not z s^k (x) $ & $\bar{s}(0)i \sigma^{\mu \nu} q_{\nu }(1- \gamma_5) b(0)$ & $g_i^{\Lambda_b \to \Lambda}$, $G_i^{\Lambda_b \to \Lambda}$ \\
  \hline
    \hline
\end{tabular}
\end{center}
 \end{table}

\subsection{Sum rules for baryonic transition  form factors}

The sum rules of form factors can be derived from the standard
procedure, that is, matching the correlation function computed in
the hadron and quark representations with the help of the dispersion
relation under the assumption of quark-hadron duality.

\subsubsection{$\Lambda_b \to p$ transition  form factors}

Inserting the complete set of states between the currents in Eq.
(\ref{correlation function}) with the same quantum numbers as
proton, we arrive at the hadronic representation of the correlator
\begin{eqnarray}
z_{\mu} T^{\mu}(P,q) = -2 f_N {(z \cdot p)^2 \over P^2 - m_N^2} [
f_1^{\Lambda_b \to p} \not z - f_2^{\Lambda_b \to p} \not z \not q
-F_1^{\Lambda_b \to p} \not z \gamma_5 + F_2^{\Lambda_b \to p} \not
z \not q \gamma_5 ] \Lambda_b (P+q) + ...\, \,\,\,
\end{eqnarray}
where the ellipsis stands for the contribution from the higher
resonance states of the proton channel and
\begin{eqnarray}
\langle 0| \epsilon^{i j k} [u^i(0) C \not z u^j (0)] \gamma_5 \not
z d^k(0)| p(P) \rangle = f_N (z \cdot P) \not z p(P)
\end{eqnarray}
have been employed in the above derivations. It needs to be pointed
out that the choice of the interpolating current for the baryon is
not unique generally. There is no general recipe to discriminate
various choices for the interpolating field of baryon.  A practical
criterion is that the coupling between the interpolating current and
the given state should be strong enough.

\begin{figure}[tb]
\begin{center}
\begin{tabular}{ccc}
\includegraphics[scale=0.6]{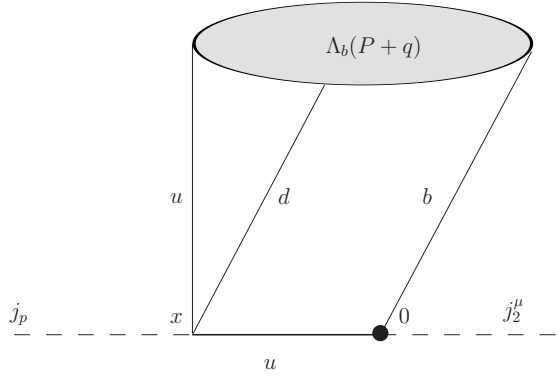}
\end{tabular}
\caption{The tree-level contribution to the correlation function
(\ref{correlation function}), where the black solid dot represent
the weak transition vertex.}\label{transition}
\end{center}
\end{figure}

On the theoretical side, the correlation function (\ref{correlation
function})  can be calculated in the perturbation theory using the
light-cone OPE. To the leading order of $\alpha_s$, the correlation
function may be computed by contracting the $u$ quark filed in
(\ref{correlation function}) and inserting the free $u$ quark
propagator
\begin{eqnarray}
z_{\mu} T^{\mu}(P,q) &=& -2 (C \not z)_{\alpha \beta} (\gamma_5 \not
z)_{\gamma^{\prime} \gamma} [\not z (1- \gamma_5)]_{\rho \tau} \int
d^4x  \int { d^4k \over (2 \pi)^4 } e^{i (p-k) \cdot x} {1 \over k^2
-m_u^2} (\not k + m_u)_{\alpha \rho} \nonumber \\
&& \times \langle 0 | \epsilon^{i j k} u^{i}_{\beta}(x)
d_{\gamma}^{j}(x) b_{\tau}^k(0) | \Lambda_b (P+q) \rangle
\label{correlator in QCD}
\end{eqnarray}
as shown in Fig. \ref{transition}.
The full quark propagator also receives corrections from the
background field \cite{asymptonic forms 1,Khodjamirian:1998ji},
which can be written as
\begin{eqnarray}
\langle 0| T \{{q_i(x) \bar{q}_j(0)}\}| 0\rangle &=& \delta_{ij}\int
{d^4 k \over (2 \pi)^4} e^{-i kx}{i \over \not \! k -m_q} -i g \int
{d^4 k \over (2 \pi)^4} e^{-i kx} \int_0^1 dv  [{1 \over 2} {\not k
+m_q \over (m_q^2 -k^2)^2} G^{\mu \nu}_{ij}(v x)\sigma_{\mu \nu }\nonumber \\
&& +{1 \over m_q^2-k^2}v x_{\mu} G^{\mu \nu}(v x)\gamma_{\nu}],
\end{eqnarray}
where the first term is the free-quark propagator and $G^{\mu
\nu}_{i j}=G_{\mu \nu}^{a} T^a_{ij}$ with ${\mbox{Tr}}[T^a T^b]={1
\over 2}\delta^{ab}$. Inserting the second term  proportional to the
gluon field strength into the correlation function can result in the
distribution amplitudes corresponding to the higher Fock states of
$\Lambda_b$ baryon. It is expected that such corrections associating
with the distribution amplitudes of higher Fock states do not play
any significant role in the sum rules for transition form factors
\cite{higher Fock state}, therefore can be neglected safely in the
presented work.

Substituting Eq. (\ref{Lambdab nonlocal matrix elements}) into Eq.
(\ref{correlator in QCD}) and performing the integral in the
coordinate space, we can achieve the correlation function in the
momentum representation at the quark level as follows:
\begin{eqnarray}
z_{\mu} T^{\mu}(P,q) &=&  2 f_{\Lambda}^{(2)} (z \cdot P)^2 \not z
(1-\gamma_5) \Lambda(v)  \int_0^{\infty} d \omega \int_0^{1} du
\bigg \{  {  m_u^2 - q^2 \over m_{\Lambda_b}^2}  { 1 \over [(P-
\omega v)^2- m_u^2 ]^2 }  \bar{\psi}_2(\omega,u) \nonumber \\
&&  + { \bar{\sigma}^2 \over [(P- \omega v)^2- m_u^2 ]^2 }
\bar{\psi}_4(\omega,u) \bigg \} + 2 f_{\Lambda}^{(1)} (z \cdot P)^2
\not z \not q (1-\gamma_5) \int_0^{\infty} d \omega \int_0^{1} du \nonumber \\
&&  { 1 \over m_{\Lambda_b} } { \bar{\sigma} \over [(P- \omega v)^2-
m_u^2 ]^2 } \bar{\psi}_3^{\sigma} (\omega,u) + ... \,
,\label{results of correlator in QCD}
\end{eqnarray}
where the subleading terms in the infinite-momentum frame kinematics
denoted by the ellipsis are suppressed. The functions
$\bar{\psi}_i(\omega,u)$ are defined as
\begin{eqnarray}
\bar{\psi}_i(\omega, u) = \int_0^{\omega} d \tau  \tau
\tilde{\psi}_{i}(\tau, u) \, ,
\end{eqnarray}
originating from the partial integral in the variable $\omega$ in
order to eliminate the factor ${1 / v \cdot x}$ due to the insertion
of distribution amplitudes in Eq. (\ref{Lambdab nonlocal matrix
elements}) \footnote{The two light-like vector $n_{\mu}$ and
$\bar{n}_{\mu}$ can be expressed by the four-velocity vector
$v_{\mu}$ and the coordinate vector $x_{\mu}$:
\begin{eqnarray}
n_{\mu} = {1 \over v \cdot x} x_{\mu}, \qquad  \bar{n}_{\mu} = 2
v_{\mu} - {1 \over v \cdot x} x_{\mu} . \nonumber
\end{eqnarray}
}.

For the convenience of matching the correlation in QCD
representation and hadronic level, Eq. (\ref{results of correlator
in QCD}) is usually written in a form of dispersion integral as
\begin{eqnarray}
z^{\mu}T_{\mu}=(z \cdot P)^2 \int_{0}^{\infty} {\rho_{V}(s, q^2)
\not \! z (1-\gamma_5) +\rho_{T}(s, q^2) \not \! z \not \! q
(1-\gamma_5) \over s-P^2}  \Lambda(v)+.... \label{dispersion
integral}
\end{eqnarray}
With the assumption of quark-hadron duality, the higher states in
the proton channel can  be  given by the same dispersion integral
only with the lower bound replaced by the effective threshold
parameter $s_0$.  Besides, the Borel transformation is commonly
introduced in the standard procedure of sum rules approach for the
sake of compensating the deficiency due to the approximation of
quark-hadron duality.

Mathematically,  the substraction of higher states can be realized
by making use of the following replacements
\begin{eqnarray}
\int_0^{\infty} d \sigma { \rho(\sigma, u) \over (P- m_{\Lambda_b}
\sigma v)^2- m_q^2 } &\rightarrow & - \int_0^{\sigma_0} { d \sigma
\over
\bar{\sigma}} { \rho(\sigma, u)  \over s-P^2}  \, , \nonumber \\
\int_0^{\infty} d \sigma { \rho(\sigma, u) \over (P- m_{\Lambda_b}
\sigma v)^2- m_q^2 }^2 &\rightarrow &  \int_0^{\sigma_0} {d \sigma
\over \bar{\sigma}^2} { \rho(\sigma, u) \over (s-P^2)^2} +{
\rho(\sigma_0, u) \over \bar{\sigma}_0^2} {1 \over s_0-P^2}
\eta(\sigma_0) , \label{replacement rule}
\end{eqnarray}
where the involved parameters are defined as
\begin{eqnarray}
\sigma= {\omega \over m_{\Lambda_b}}, \hspace{ 1 cm }
\bar{\sigma}=1-\sigma, \hspace{ 1 cm } s= \sigma m_{\Lambda_b}^2 + {
m_q^2- \sigma q^2 \over \bar{\sigma} }, \hspace{ 1 cm }
\eta(\sigma_0) = (1+ { m_q^2 -q^2 \over \bar{\sigma}^2
m_{\Lambda_b}^2 } )^{-1} ,
\end{eqnarray}
and $\sigma_0$ is the positive solution of the corresponding
quadratic equation for $s=s_0$:
\begin{eqnarray}
\sigma_0={(s_0+m_{\Lambda_b}^2-q^2) -
\sqrt{(s_0+m_{\Lambda_b}^2-q^2)^2-4m_{\Lambda_b}^2(s_0-m_q^2)}\over
2M^2} .
\end{eqnarray}

Performing the Borel transformation, we can finally obtain the sum
rules for the transition form factors
\begin{eqnarray}
f_N f_1^{\Lambda_b \to p}(q^2) e^{- {m_N^2 \over M^2} } &=&
f_{\Lambda}^{(2)} \int_0^1 du \bigg \{  { m_{\Lambda_b} \over M^2}
\int_0^{\sigma_0} {d \sigma \over \bar{\sigma}^2} \bigg[ {m_u^2 -q^2
\over m_{\Lambda_b}^2  } \bar{\psi}_2(\omega,u) + \bar{\sigma}^2
\bar{\psi}_4(\omega,u) \bigg] e^{-s/M^2}
\nonumber \\
&& + {1 \over m_{\Lambda_b} \bar{\sigma}_0^2 } \bigg [ {m_u^2 -q^2
\over m_{\Lambda_b}^2  } \bar{\psi}_2(\omega_0,u) + \bar{\sigma}_0^2
\bar{\psi}_4(\omega_0,u)
\bigg ] \eta(\sigma_0)  e^{-s_0^{p} /M^2} \bigg \} , \label{sum rules for lambdab to proton} \\
f_N f_2^{\Lambda_b \to p}(q^2) e^{- {m_N^2 \over M^2} } &=& -
f_{\Lambda}^{(1)} \int_0^1 du \bigg \{  { 1 \over M^2 }
\int_0^{\sigma_0} { d \sigma \over \bar{\sigma} }
\bar{\psi}_3^{\sigma} ( \omega,u) e^{-s /M^2} + {1 \over
m_{\Lambda_b}^2 }  { 1 \over \bar{\sigma}_0} \bar{\psi}_3^{\sigma} (
\omega_0,u) \eta(\sigma_0) e^{-s_0^{p} /M^2}  \bigg \} , \nonumber
\end{eqnarray}
with $\omega_0 = m_{\Lambda_b} \sigma_0$ and $s_0^{p}$ being the
threshold value of proton channel. It may be easily observed that
the sum rules for the form factors $F_1^{\Lambda_b \to p}(q^2)$ and
$F_2^{\Lambda_b \to p}(q^2)$ are the same as those for
$f_1^{\Lambda_b \to p}(q^2)$ and $f_2^{\Lambda_b \to p}(q^2)$
respectively
\begin{eqnarray}
F_1^{\Lambda_b \to p}(q^2)=f_1^{\Lambda_b \to p}(q^2),  \qquad
F_2^{\Lambda_b \to p}(q^2)=f_2^{\Lambda_b \to p}(q^2).
\end{eqnarray}

\subsubsection{$\Lambda_b \to \Lambda$ transition  form factors}

Following the same procedure, we can achieve the hadronic
representation of the correlator responsible for the tensor
$\Lambda_b \to \Lambda$  transition
\begin{eqnarray}
z_{\mu} T^{\mu}(P,q) = -2 f_{\Lambda} {(z \cdot p)^2 \over P^2 -
m_{\Lambda}^2} [ g_1^{\Lambda_b \to \Lambda} \not z - g_2^{\Lambda_b
\to \Lambda} \not z \not q + G_1^{\Lambda_b \to \Lambda} \not z
\gamma_5 - G_2^{\Lambda_b \to \Lambda} \not z \not q \gamma_5 ]
\Lambda_b (P+q) + ......\,
\end{eqnarray}
where the ellipsis stands for the contribution from the higher
resonance states of $\Lambda$ baryon channel. The coupling between
the selected current and $\Lambda$ baryon is given by
\begin{eqnarray}
\langle 0| \epsilon^{i j k} [u^i(0) C \gamma_5 \not z d^j (0)] \not
z s^k(0)| \Lambda(P) \rangle = f_{\Lambda} (z \cdot P) \not z
\Lambda(P) .
\end{eqnarray}

Similarly, one can also obtain the QCD representation of the
correlation function in terms of the light-cone OPE
\begin{eqnarray}
z_{\mu} T^{\mu}(P,q) &=& 2 f_{\Lambda}^{(2)} (z \cdot P)^2 \not z
\not q (1+ \gamma_5) \Lambda(v)   \nonumber \\ && \times
\int_0^{\infty} d \omega \int_0^1 du { 1 \over [(P- \omega v)^2-
m_s^2 ]^2 } \bigg [ { q^2 -m_s^2 \over m_{\Lambda_b}^2}
\bar{\psi}_2(\omega, u) - \bar{\sigma}^2 \bar{\psi}_4(\omega, u)
\bigg ]  + ...\, ,
\end{eqnarray}
where the subleading terms in the infinite-momentum frame kinematics
represented by the ellipsis are neglected. It is obvious that the
form factors $g_1^{\Lambda_b \to \Lambda}(q^2)$ and $G_1^{\Lambda_b
\to \Lambda}(q^2)$ do not contribute to the correlation function
associated with the $\Lambda_b \to \Lambda$  transition at the
leading power.

Matching the correlation function in the above two representations
and performing the Borel transformation with the variable $P^2$, one
can arrive at the sum rules for the transition form factors
\begin{eqnarray}
 f_{\Lambda} g_2^{\Lambda_b \to \Lambda}(q^2) e^{- {m_{\Lambda}^2
\over M^2} } &=& f_{\Lambda}^{(2)} \int_0^1 du \bigg \{  {
m_{\Lambda_b} \over M^2 }
 \int_0^{\sigma_0} d \sigma
[{m_s^2 -q^2 \over m_{\Lambda_b}^2} \bar{\psi}_2(\omega,u)  +
\bar{\sigma}^2 \bar{\psi}_4(\omega,u) ] e^{-s /M^2}  \nonumber \\
&& +  { 1 \over m_{\Lambda_b} \bar{\sigma}_0^2} [{m_s^2 -q^2 \over
m_{\Lambda_b}^2} \bar{\psi}_2(\omega_0,u)  + \bar{\sigma}_0^2
\bar{\psi}_4(\omega_0,u) ] \eta(\sigma_0) e^{-s_0^{\Lambda} /M^2}
\bigg \}, \label{sum rules for lambdab to lambda}
\end{eqnarray}
where $s_0^{\Lambda}$ is the duality-threshold parameter of
$\Lambda$ channel. In addition, the sum rules for the form factors
satisfy
\begin{eqnarray}
f_1^{\Lambda_b \to \Lambda}(q^2)=F_1^{\Lambda_b \to
\Lambda}(q^2)=G_2^{\Lambda_b \to \Lambda}(q^2)=g_2^{\Lambda_b \to
\Lambda}(q^2) .
\end{eqnarray}

\section{Numerical analysis of sum rules for form factors}

Now, we are going to calculate the form factors $f_1^{\Lambda_b \to
p}(q^2)$, $f_2^{\Lambda_b \to p}(q^2)$ responsible for the
$\Lambda_b \to p$ decay and $g_2^{\Lambda_b \to \Lambda}(q^2) $
relevant to the $\Lambda_b \to \Lambda$ transition numerically.
Firstly, we collect the input parameters used in this
paper\cite{Ball:2008fw,Amsler:2008zzb,ioffe,Braun:2006hz,Liu:2008yg}
\begin{equation}
\begin{array}{ll}
m_s(1 {\rm{GeV}})=142 {\rm{MeV}}, &m_u(1 {\rm{GeV}})=2.8 {\rm{MeV}},
\\
m_{\Lambda_b}=5.62 {\rm{GeV}}, &  m_{p}=0.938 {\rm{GeV}},
\\
m_{\Lambda}=1.12 {\rm{GeV}}, &
f_{\Lambda}^{(1)}=f_{\Lambda}^{(2)}=0.030 \pm 0.005 {\rm{GeV}}^{3}
\\
f_N=(5.0 \pm 0.5)\times 10^{-3} {\rm{GeV}^2}, & f_{\Lambda}=(6.0 \pm
0.3)\times 10^{-3} {\rm{GeV}^2}
\\
s_0^{p}=(2.25 \pm 0.10) {\rm{GeV}}^2, & s_0^{\Lambda}=(2.55 \pm
0.10) {\rm{GeV}}^2.\label{inputs}
\end{array}
\end{equation}
The normalization constants of the light-cone distribution
amplitudes for the proton, $\Lambda$ baryon, and $\Lambda_b$ baryon,
namely, $f_N$, $f_{\Lambda}$, $f_{\Lambda}^{(1)}$ and
$f_{\Lambda}^{(2)}$ are all evaluated at the renormalization scale
$\mu= 1{\rm  GeV}$. As for the choice of the threshold parameters
$s_0^{p}$ and $s_0^{\Lambda}$, one should determine it by demanding
the sum rules to be relatively stable in allowed regions for Borel
mass $M_B^2$, the value of which should be around the mass square of
the corresponding first excited states. As for the heavy-light
systems, the standard value of the threshold in the $X$ channel
would be $s_0^{X}=(m_X+\Delta_X)^2$, where $\Delta_X$ is about $0.5$
GeV \cite{dosch, matheus, bracco, navarra, Wang:2007ys, Colangelo}.

With   all the parameters listed above, we can proceed to compute
the numerical values of the form factors. The form factors should
not depend on the the Borel mass $M^2$ in a complete theory.
However, as we truncate the  operator product expansion up to
next-to-leading conformal spin for the $\Lambda_b$ bayon in the
leading Fock configuration and keep the perturbative expansion in
$\alpha_s$ to leading order, a manifest dependence of the form
factors on the Borel parameter $M^2$ would emerge in practice.
Therefore, one should look for a working ``window", where the
results only mildly vary with respect to the Borel mass, so that the
truncation is acceptable.

\begin{figure}
\begin{center}
\includegraphics[width=7.0 cm]{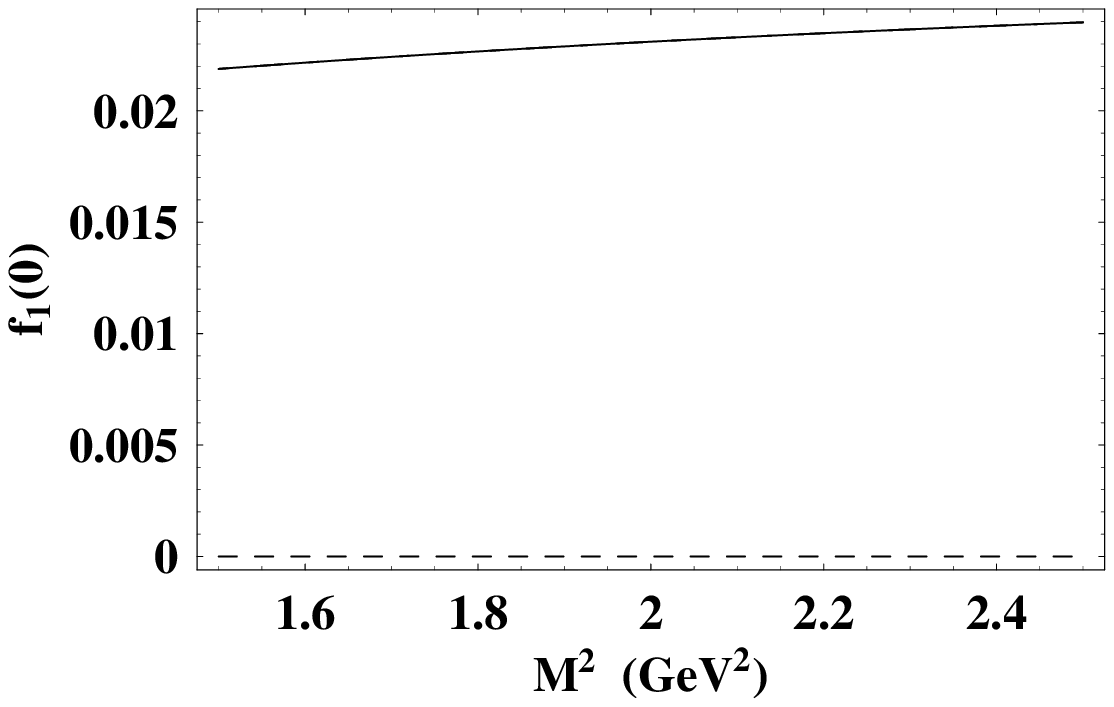}
\includegraphics[width=7.0 cm]{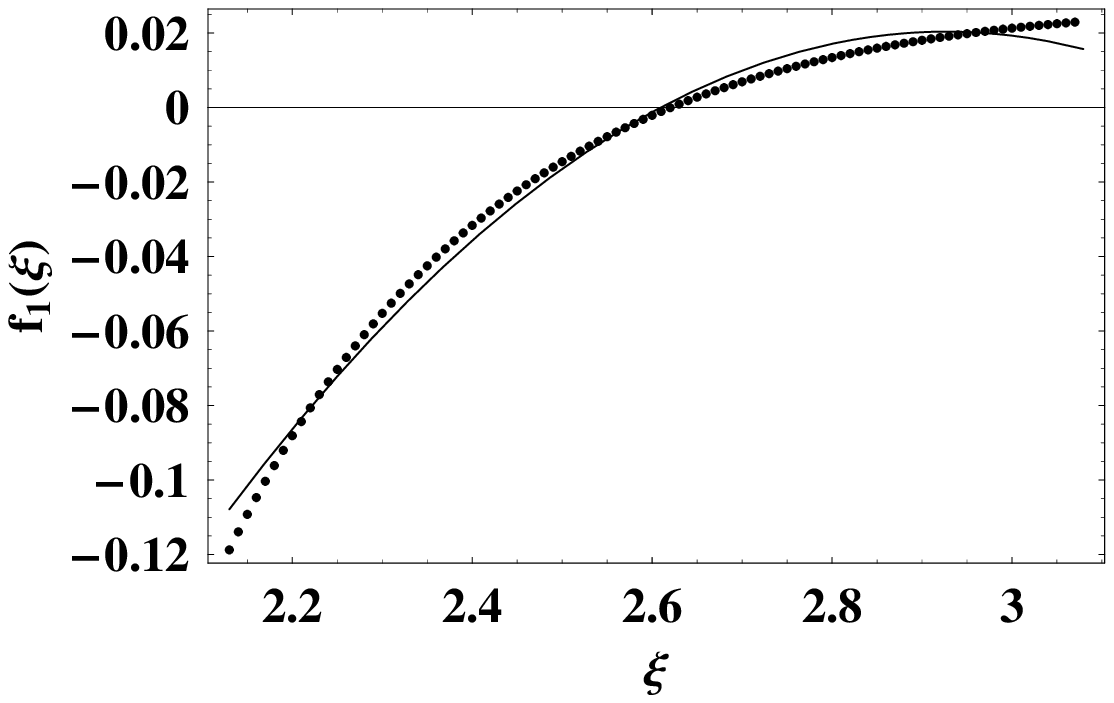}
\vspace{-1.5 cm}
 \caption{Left panel: The dependence of form factor
$f_1^{\Lambda_b \to p}(q^2=0)$ on the Borel mass $M^2$(solid line)
and the contribution from the twist-2 distribution amplitude in  the
whole sum rules (dashed line); Right panel: The dependence of form
factor $f_1^{\Lambda_b \to p}$ on the momentum transfer $q^2$ within
the kinematical region, where the light-cone OPE for the correlation
function works well. The dashed line represents the result of form
factor $f_1^{\Lambda_b \to p}$ predicted by the $\Lambda_b$-baryon
LCSR, while the solid line describes the form factor $f_1^{\Lambda_b
\to p}$ determined by  Eq. (\ref{parameterization of form
factors}).}
 \label{fig:form factor proton f1}
 \end{center}
 \end{figure}

\begin{figure}
\begin{center}
\includegraphics[width=7.0 cm]{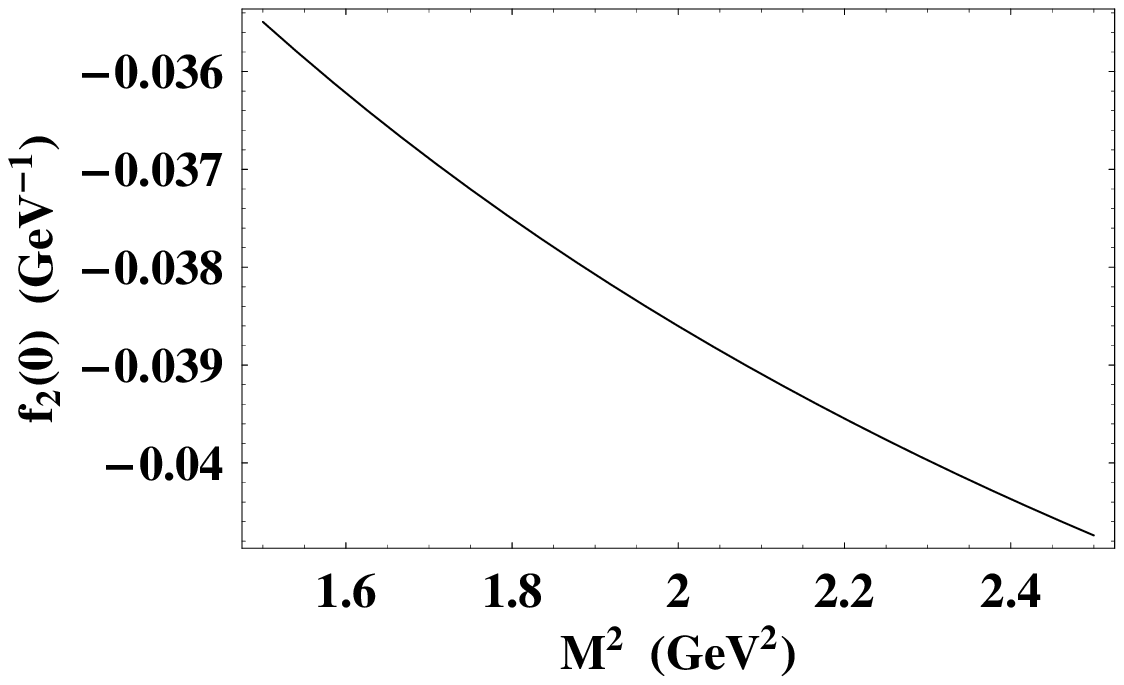}
\includegraphics[width=7.0 cm]{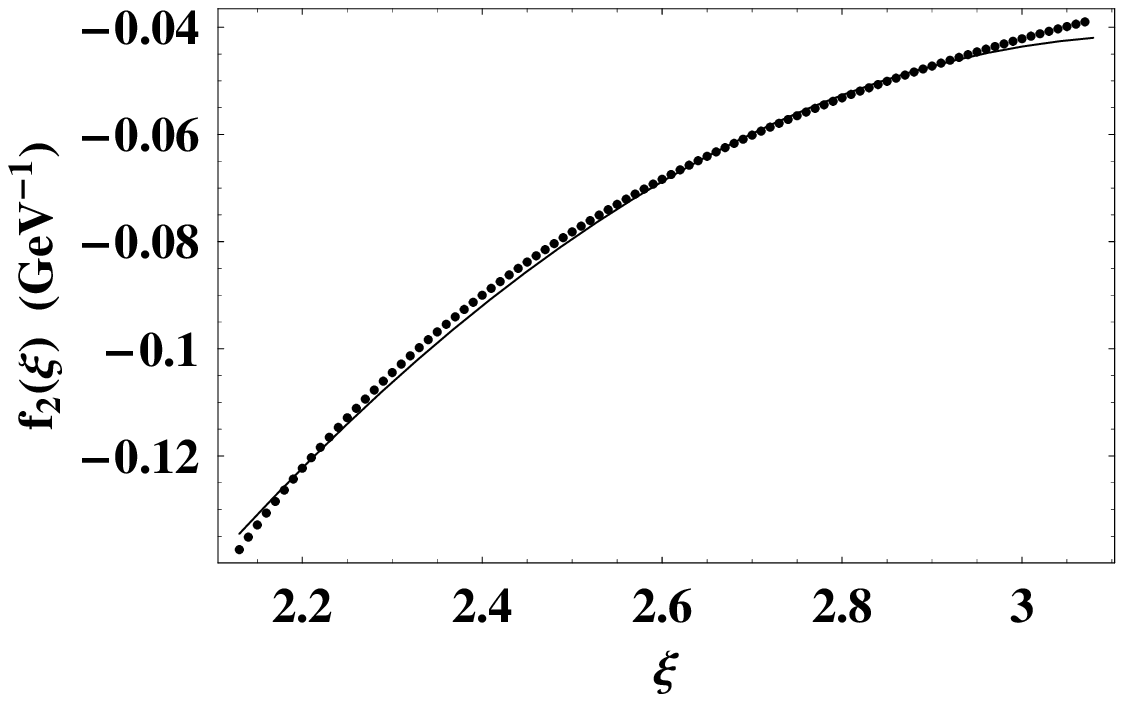}
\vspace{-1.5 cm}
 \caption{Left panel: The dependence of form factor
$f_2^{\Lambda_b \to p}(q^2=0)$ on the Borel mass $M^2$; Right panel:
The dependence of form factor $f_2^{\Lambda_b \to p}$ on the
momentum transfer $q^2$ within the kinematical region, where the
light-cone OPE for the correlation function is valid. The dashed
line represents the result of form factor $f_2^{\Lambda_b \to p}$
predicted by the $\Lambda_b$-baryon LCSR, while the solid line
describes the form factor $f_2^{\Lambda_b \to p}$ determined by Eq.
(\ref{parameterization of form factors}).}
 \label{fig:form factor proton f2}
 \end{center}
 \end{figure}

Firstly, we focus on the form factors at the zero-momentum transfer.
As shown in Fig. \ref{fig:form factor proton f1},  the form factor
$f_1^{\Lambda_b \to p}$ is rather stable with the selected Borel
mass $M^2 \in [1.5, 2.5] {\rm GeV^2}$, which is consistent with that
determined from the two-point sum rules for the nucleon form factors
\cite{Braun:2001tj}. In principle,  the Borel parameter $M^2$ should
not be too large in order to insure that the contributions from the
higher states are exponentially damped as can be observed form Eq.
(\ref{sum rules for lambdab to proton}) and the global quark-hadron
duality is satisfactory. On the other hand, the Borel mass   could
not be too small for the validity of OPE near the light-cone for the
correlation function in deep Euclidean region, since the
contributions of higher twist distribution amplitudes amount the
higher order of ${1 / M^2}$ to the leading contributions.
 The value of $f_1^{\Lambda_b \to p}(q^2=0)$ is $0.023^{+0.006}_{-0.005}$, where the
uncertainties from the variations of Borel parameters, the
fluctuation of the threshold value and the uncertainties from the
normalization constants of  the hadronic distribution amplitudes are
combined together. Following the same procedure, we can continue to
estimate the numerical results for the form factor $f_2^{\Lambda_b
\to p}(q^2=0)$ at the zero-momentum transfer within the chosen Borel
window as displayed in Fig. \ref{fig:form factor proton f2}. It may
be observed that $f_2^{\Lambda_b \to p}(q^2=0) =
-0.039^{+0.009}_{-0.009} {\rm GeV^{-1}} $ with the given Borel
window $M^2 \in [1.5, 2.5] {\rm GeV^2}$.
As for the $\Lambda_b \to \Lambda$ transition, the Borel platform
for the form factor $g_2^{\Lambda_b \to \Lambda}$ at the
zero-momentum transfer is determined as $M^2 \in [2.0, 3.0] {\rm
GeV^2}$ with the threshold parameter $s_0^{\Lambda}=2.55
{\rm{GeV}}^2$. It can be observed from Fig. \ref{fig:form factor
lambda g2} that $g_2^{\Lambda_b \to
\Lambda}(q^2=0)=0.018^{+0.003}_{-0.003} $. To illustrate the
SU(3)-breaking effects predicted in the $\Lambda_b$-baryon LCSR, it
is helpful to define the following ratio
\begin{eqnarray}
R \equiv { f_1^{\Lambda_b \to p}(q^2=0)  \over f_1^{\Lambda_b \to
\Lambda}(q^2=0)} =  1.28^{+0.14}_{-0.08}.
\end{eqnarray}
In particular, this ratio is less sensitive to the variations of
hadronic  parameters involved in the $\Lambda_b$ baryon distribution
amplitudes than the individual form factors.  It can be easily
observed that SU(3) violating effects are attributed to the
differences between the masses of $s$ and $u$ quarks and the
discrepancy in the duality-threshold parameters for proton and
$\Lambda$ baryon.
\begin{figure}
\begin{center}
\includegraphics[width=7.0 cm]{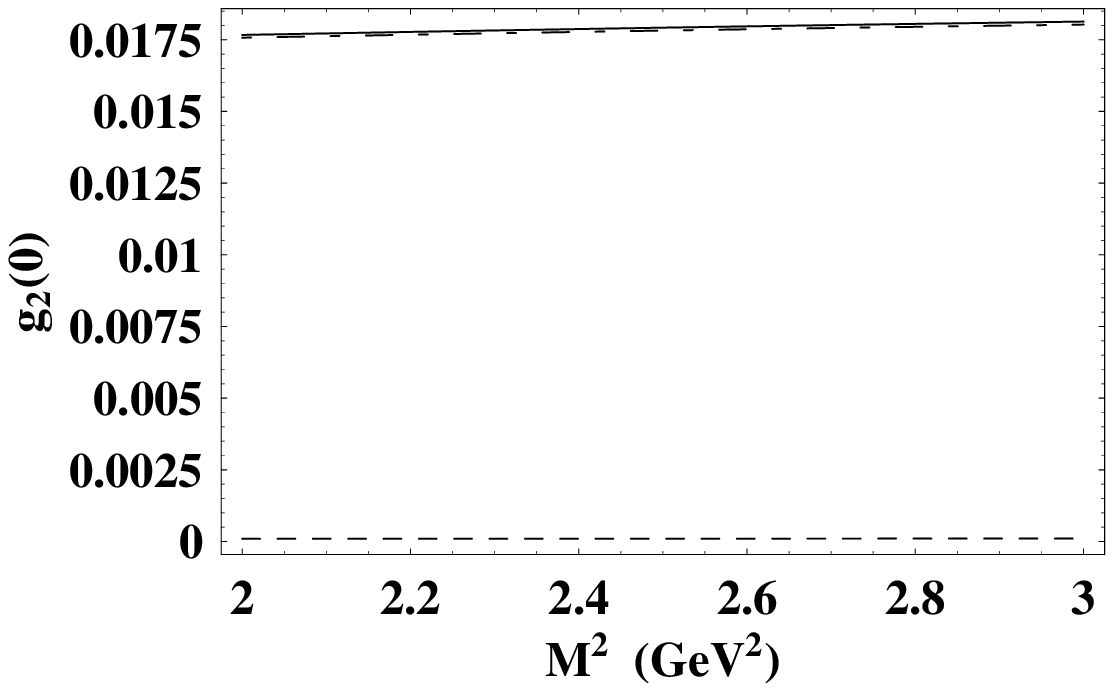}
\includegraphics[width=7.0 cm]{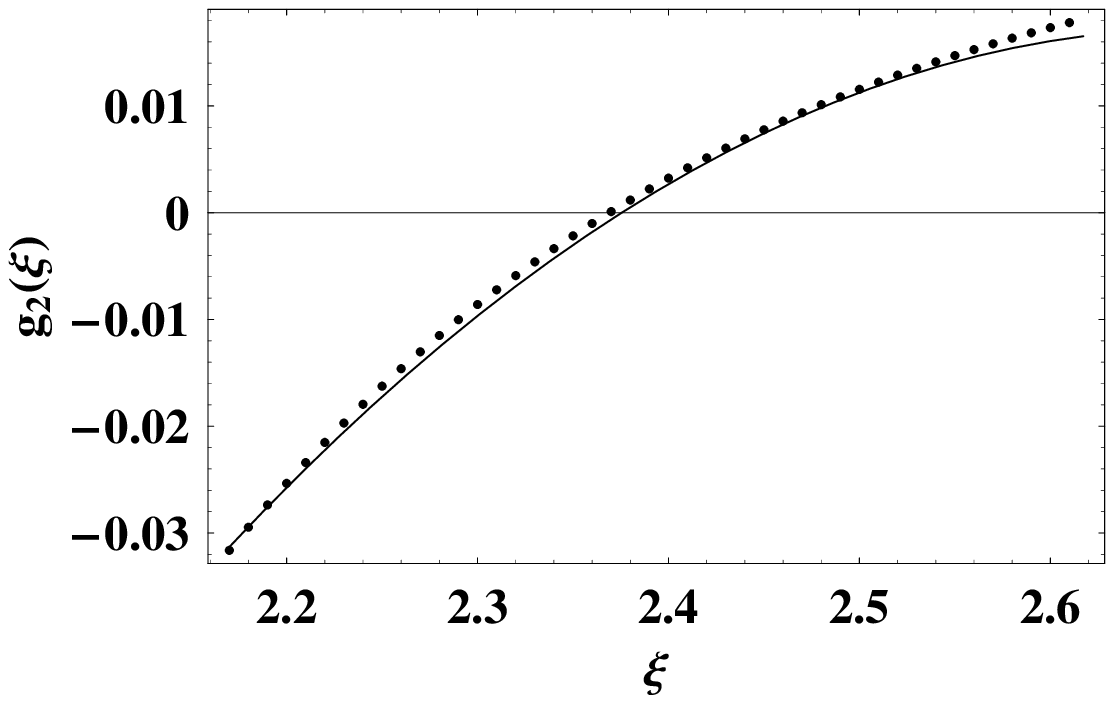}
\vspace{-1.5 cm}
 \caption{Left panel: The dependence of form factor
$g_2^{\Lambda_b \to \Lambda}(q^2=0)$ on the Borel mass $M^2$(solid
line) and the contribution from the twist-2 distribution amplitude
in  the whole sum rules (dashed line); Right panel: The dependence
of form factor $g_2^{\Lambda_b \to \Lambda}$ on the momentum
transfer $q^2$ within the kinematical region, where the light-cone
OPE for the correlation function works well. The dashed line
represents the result of form factor $g_2^{\Lambda_b \to \Lambda}$
predicted by the $\Lambda_b$-baryon LCSR, while the solid line
describes the form factor $g_2^{\Lambda_b \to \Lambda}$ determined
by  Eq. (\ref{parameterization of form factors}).}
 \label{fig:form factor lambda g2}
 \end{center}
 \end{figure}

In Table \ref{results of form factors at zero momentum transfer},
the numbers of the various transition form factors predicted in the
$\Lambda_b$-baryon LCSR, light-baryon LCSR
\cite{Huang:2004vf,Wang:2008sm}, three-point QCDSR
\cite{Huang:1998ek,Huang:1998rq} as well as PQCD approach
\cite{Lu:2009cm,He:2006ud} are grouped together. It is worthwhile to
remind that the quark-hadron duality is employed differently in the
$\Lambda_b$-baryon and light-baryon LCSR. Consequently, the
difference between the predictions of two kinds of LCSR for the same
transition form factors can be taken into account as a quantitative
estimation of the systematic uncertainty. As shown in this table,
the $\Lambda_b \to p$ transition form factors evaluated in
$\Lambda_b$-baryon LCSR and light-baryon LCSR with full QCD are
basically consistent with each other, which implies that the power
corrections to the $\Lambda_b$-baryon LCSR are numerically small.
The predictions of $\Lambda_b \to p$ transition form factors in
terms of the light-baryon LCSR with HQET deviate distinctly from
that obtained in the light-baryon LCSR with full QCD. Different
versions of LCSR can be easily discriminated by measureming the
semeleptonic $\Lambda_b \to p$ decay at the ongoing and forthcoming
colliders. The $\Lambda_b \to \Lambda$ transition form factor
$g_2^{\Lambda_b \to \Lambda}(q^2=0)$ estimated in light-baryon LCSR
is almost one order larger than that given by the $\Lambda_b$-baryon
LCSR. Such distinct discrepancy can be attributed to the fact that
only the asymptotic contributions of distribution amplitudes for the
$\Lambda$ baryon are included in the sum rules for the form factor
$g_2^{\Lambda_b \to \Lambda}$, apart form the systematic
uncertainties coming from the different quark-hadron duality
assumptions as mentioned above.

It is also observed from Table \ref{results of form factors at zero
momentum transfer} that the hard contributions to the form factor
$f_{1}^{\Lambda_b \to p}$ involving two hard-gluons' exchange,
estimated in the PQCD approach, is  approximately one order smaller
than those contributions dominated by the soft gluon exchange as
estimated in $\Lambda_b$-baryon  LCSR. In other words, the
$\Lambda_b \to p$ transition form factors are dominated by the
non-perturbative contributions, which may not be estimated reliably
in the PQCD approach. As for the $\Lambda_b \to \Lambda$ transition,
the value of form factor $f_1^{\Lambda_b \to \Lambda}$
(=$g_2^{\Lambda_b \to \Lambda}$) is approximately $5$ times larger
than that of $f_1^{\Lambda_b \to \Lambda}$ in PQCD approach,
implying unexpectable SU(3) symmetry breaking effects. A potential
reason responsible for such impenetrable observation  is that the
distribution amplitudes of $\Lambda$ baryon employed in the analysis
of Ref. \cite{He:2006ud} are motivated by quark model, which are not
consistent with the QCD constraints.
\begin{table}
 \caption{$\Lambda_b \to p$ and $\Lambda_b \to \Lambda$
 transition form factors computed in the LCSR with $\Lambda_b$
 distribution amplitudes, where the uncertainties from
 the Borel mass, threshold parameter and normalization constants
 of the hadronic distribution amplitudes are combined together.
For comparison, we also cite the theoretical
 estimations of the form factors in the LCSR with light baryon
 distribution amplitudes, three-point QCD sum rules  and PQCD approach.}
 \label{results of form factors at zero momentum transfer}
 \begin{center}
\begin{tabular}{c|ccccc}
  \hline
  \hline
  form factor & $\Lambda_b$ baryon LCSR & light-baryon LCSR  & light-baryon LCSR  & 3-point QCDSR & PQCD\\
  &   &   HQET &  full QCD  &      &\\
  $f_1^{\Lambda_b \to p}(q^2=0)$ & $0.023^{+0.006}_{-0.005}$ &  $-2.14 \times 10^{-3}$ \cite{Huang:2004vf} &  $0.018$ \cite{Huang:2004vf}  & $0.22$ \cite{Huang:1998rq} & $2.2^{+0.8}_{-0.5} \times 10^{-3}$ \cite{Lu:2009cm} \\
  $f_2^{\Lambda_b \to p}(q^2=0) \, (\rm GeV^{-1}) $  & $- 0.039^{+0.009}_{-0.009}$  & $- 0.015$ \cite{Huang:2004vf} &  $-0.028$ \cite{Huang:2004vf} & $0.71 \times 10^{-2}$\cite{Huang:1998rq} &  --- \\
  $g_2^{\Lambda_b \to \Lambda}(q^2=0)$& $0.018^{+0.003}_{-0.003}$ &  --- &   $0.14^{+0.02}_{-0.01}$ \cite{Wang:2008sm} & $0.45$ \cite{Huang:1998ek} &  $(1.2-1.6) \times 10^{-2}$  \cite{He:2006ud} \\
  \hline
  \hline
\end{tabular}
\end{center}
 \end{table}

In the next place, we can further investigate the $q^2$ dependence
of the $\Lambda_b \to p, \Lambda$ form factors basd on the sum rules
given in  (\ref{sum rules for lambdab to proton}) and (\ref{sum
rules for lambdab to lambda}). As already mentioned, the light-cone
OPE is expected to be successful at the region (\ref{region for
OPE}), which indicates that the sum rules for  transition form
factors are reliable only for $0 < q^2  < 10 \,\, {\rm GeV^2}$.
 LCSR  with the light-baryon distribution amplitudes are applicable
at larger momentum transfer, up to $14-16 {\rm GeV^2}$
\cite{Wang:2008sm}. The dependence of  form factors $f_1^{\Lambda_b
\to p}$, $f_2^{\Lambda_b \to p}$ and $g_2^{\Lambda_b \to \Lambda}$
on the momentum transfer  have been plotted in Fig. (\ref{fig:form
factor proton f1}), (\ref{fig:form factor proton f2}) and
(\ref{fig:form factor lambda g2}) respectively.

\begin{figure}
\begin{center}
\includegraphics[width=7.0 cm]{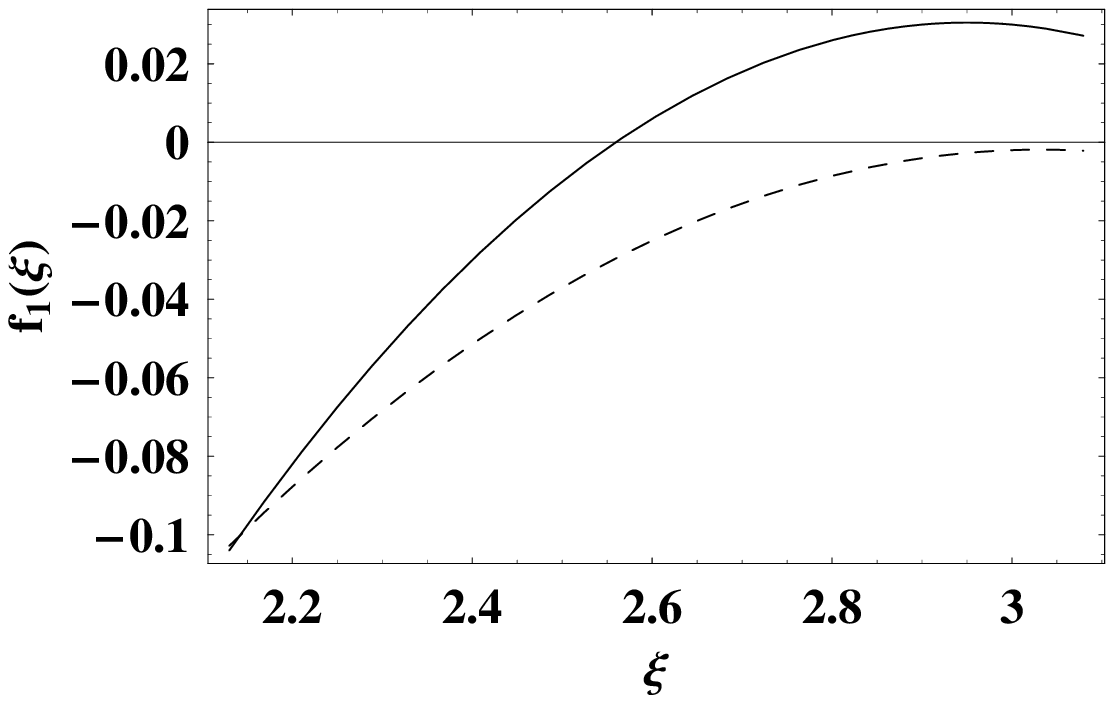}
\includegraphics[width=7.0 cm]{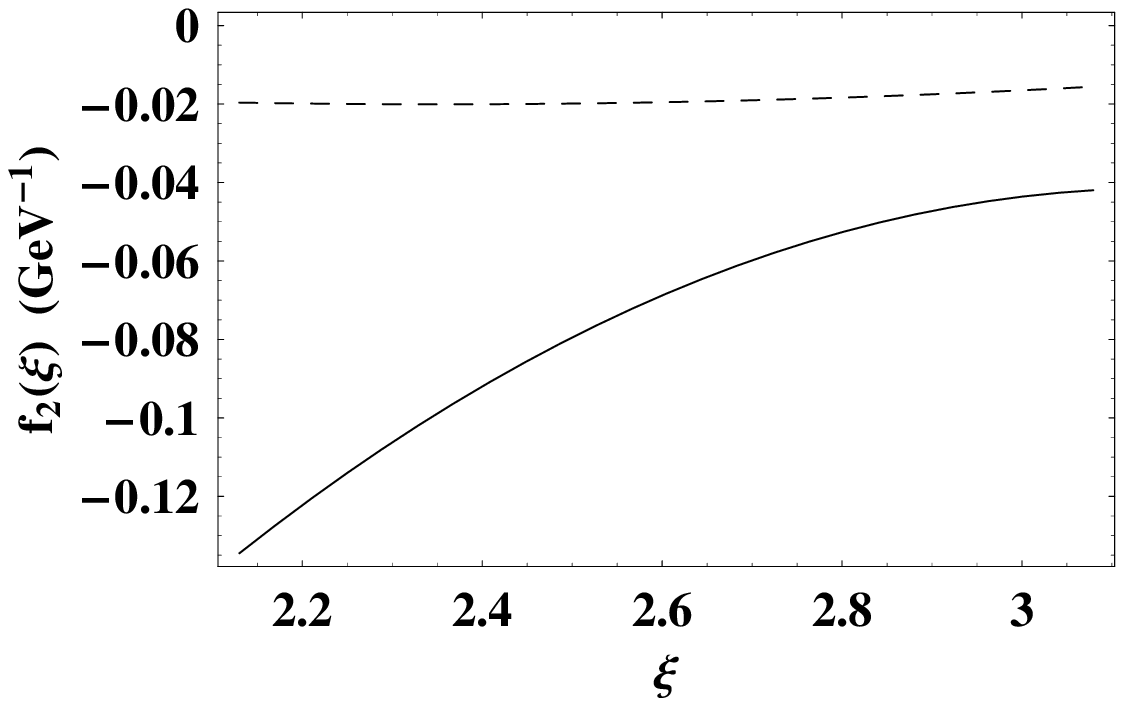}
\vspace{-1.5 cm}
 \caption{The dependence of form factors $f_1^{\Lambda_b \to p}$ and  $f_2^{\Lambda_b \to p}$
 on the momentum transfer evaluated in two LCSR: the solid line denotes the
 results given by the $\Lambda_b$-baryon LCSR, while the dashed line
 describes the predictions from the light-baryon LCSR.}
 \label{fig:form factors comparision}
 \end{center}
 \end{figure}

Following Ref. \cite{Huang:2004vf}, we fit the results of form
factors given by the $\Lambda_b$-LCSR at $0 < q^2  < 10 {\rm GeV^2}$
to the following parameterization
\begin{eqnarray}
\eta_i (\xi) = a_i + b_i \xi + c_i \xi, \label{parameterization of
form factors}
\end{eqnarray}
where the label $\xi_i$ denote the form factors  $f_1^{\Lambda_b \to
p}$, $f_2^{\Lambda_b \to p}$ and $g_2^{\Lambda_b \to \Lambda}$. The
 viable $\xi$ is defined as $\xi={v \cdot P \over m_{L}}$, whose
 number in the whole kinematical region is
 $1 \leq \xi \leq \xi_{\rm max}\equiv {m_{\Lambda_b}^2 + m_L^2 \over m_{\Lambda_b}
 m_L}$. The values of parameters $a_i$, $b_i$ and $c_i$ are tabulated in
Table \ref{results of form factors with momentum transfer}.
\begin{table}
\caption{The parameters $a_i$, $b_i$ and $c_i$ in Eq.
(\ref{parameterization of form factors}) determined by the
$\Lambda_b$-baryon LCSR at  the region $0 < q^2  < 10 {\rm GeV^2}$,
where the uncertainties from
 the Borel mass, threshold parameter and normalization constants
 of the hadronic distribution amplitudes are combined together.
The results obtained in the light-baryon LCSR \cite{Huang:2004vf}
are also collected for comparison.}
 \label{results of form factors with momentum transfer}
 \begin{center}
\begin{tabular}{c|ccc}
  \hline
  \hline
  form factor & \hspace{2 cm} $a_i$ & \hspace{2 cm}  $b_i$ & \hspace{2 cm}  $c_i$ \\
  $f_1^{\Lambda_b \to p}$& \hspace{2 cm}  $-1.71^{+0.44}_{-0.42}$ & \hspace{2 cm}  $1.18^{+0.28}_{-0.30}$ & \hspace{2 cm}  $-0.20^{+0.05}_{-0.05}$ \\
  & \hspace{1 cm} $-1.14$ \cite{Huang:2004vf} & \hspace{2 cm} $0.75$ \cite{Huang:2004vf} & \hspace{2 cm} $-0.12$ \cite{Huang:2004vf} \\
  $f_2^{\Lambda_b \to p} \, (\rm GeV^{-1})$ & \hspace{2 cm}  $-0.92^{+0.18}_{-0.19}$ & \hspace{2 cm}  $0.56^{+0.11}_{-0.11}$ & \hspace{2 cm}  $-0.088^{+0.018}_{-0.018}$ \\
  & \hspace{2 cm} $0.027$ \cite{Huang:2004vf} & \hspace{2 cm} $-0.040$ \cite{Huang:2004vf} & \hspace{2 cm} $0.0085$ \cite{Huang:2004vf} \\
  $g_2^{\Lambda_b \to \Lambda}$ & \hspace{2 cm}  $-1.33^{+0.26}_{-0.26}$ & \hspace{2 cm}  $1.01^{+0.20}_{-0.20}$ & \hspace{2 cm}  $-0.19^{+0.04}_{-0.03}$ \\
  \hline
  \hline
\end{tabular}
\end{center}
 \end{table}
It is clear from this table that two different LCSR can lead to the
consistent numbers  for the $q^2$-dependence of the form factor
$f_1^{\Lambda_b \to p}$, which have been displayed in Fig.
\ref{fig:form factors comparision}. However, the form factor
$f_2^{\Lambda_b \to p}$ estimated in the $\Lambda_b$-baryon LCSR
deviates from that predicted in the light-baryon LCSR significantly,
implying that the power corrections in these two LCSR differ from
each other remarkably.  To illustrate the discrepancy more
quantitatively, we present the $q^2$-dependent behavior of the form
factor $f_2^{\Lambda_b \to p}$ in two LCSR manifestly.
As shown in Fig. \ref{fig:form factors comparision}, the form factor
$f_2^{\Lambda_b \to p}$ rises drastically with increasing squared
momentum transfer $q^2$ in the $\Lambda_b$-baryon LCSR, however, it
almost does not change with the variation of $q^2$ in the
light-baryon LCSR. 

\section{Discussions and conclusions}

Employing the distribution amplitudes of $\Lambda_b$ baryon, we
investigate the $\Lambda_b \to p, \,\, \Lambda$ transition form
factors in the framework of LCSR approach, which serves as a further
development of $B$-meson LCSR suggested in Ref.
\cite{Khodjamirian:2005ea,Khodjamirian:2006st}. Such transition form
factors play the role of a corner stone to explore the quark-flavor
structure of the SM as well as determine its fundamental parameters
such as the CKM matrix. Pinning down the uncertainties of transition
form factors in many cases is an essential prescription to improve
the accuracy of theoretical predictions.

Our results indicate that $\Lambda_b \to p$ transition form factors
computed in the $\Lambda_b$-baryon LCSR are consistent with that
given by the standard LCSR in full QCD within the error bars. Such
agreement possibly illustrate that the power corrections to the
$\Lambda_b$-baryon LCSR are not sizable. The prediction of form
factor $g_2^{\Lambda_b \to \Lambda}(q^2=0)$ in light-baryon LCSR is
about one order larger than that estimated in the $\Lambda_b$-baryon
LCSR. It is not difficult to understand these results from two kinds
of LCSR. In the $\Lambda_b$-baryon LCSR, the distribution amplitudes
of $\Lambda_b$ baryon are the universal nonperturbative inputs,
which parameterize the long distance dynamics of $\Lambda_b \to p,
\, \Lambda$ form factors. However, it is the distribution amplitudes
of light baryon that describe the nonperturbative contributions to
the form factors in the standard LCSR. It is well known that
hadronic distribution amplitudes are of limited accuracy due to our
lack good understanding of QCD at low energies. Reasonable
prediction on the SU(3) symmetry breaking effects between the form
factors $f_1^{\Lambda_b \to p}$ and $f_1^{\Lambda_b \to \Lambda}$ in
the $\Lambda_b$-baryon LCSR can be ascribed to the same hadronic
distribution amplitudes involved in the sum rules, which can reduce
the theoretical uncertainties significantly. As for the light baryon
LCSR, the distribution amplitudes of proton are considered up to the
next-to-leading conformal spin accuracy in Ref. \cite{Huang:2004vf};
while only the asymptotic forms of $\Lambda$ baryon distribution
amplitudes are included in the sum rules for the form factor
$g_2^{\Lambda_b \to \Lambda}$ in Ref. \cite{Wang:2008sm}. Hence, it
is probable that the pre-asymptonic corrections to the $\Lambda$
baryon are quite crucial to reconcile the existing discrepancy. As a
matter of fact, large pre-asymptonic corrections to the distribution
amplitudes of proton have been observed in the form factors
responsible for  $\Lambda_b \to p$ transition in Ref.
\cite{Huang:2004vf}.

Within the framework of $\Lambda_b$-baryon LCSR, we also study the
dependence of form factors on the momentum transfer $q^2$. It is
shown that the $\Lambda_b$-baryon LCSR prediction of $f_1^{\Lambda_b
\to p}$ is in accord with that estimated in the light-baryon LCSR.
Moreover, radiative corrections to the $\Lambda_b \to p, \, \Lambda$
transition form factors can be further carried out, once the
renormalization-group evolutions for the $\Lambda_b$ distribution
amplitudes are available.

\section*{Acknowledgments}

This work is partly supported by the National Science Foundation of
China under Grant No.10735080 and 10625525. The authors would like
to thank Vladimir Braun for helpful discussions.



\begin{thebibliography}{9}





\bibitem{Li:1992ce}
  H.~n.~Li,
  Phys.\ Rev.\  D {\bf 48}, 4243 (1993).


\bibitem{Kundu:1998gv}
  B.~Kundu, H.~n.~Li, J.~Samuelsson and P.~Jain,
  Eur.\ Phys.\ J.\  C {\bf 8}, 637 (1999)
  [arXiv:hep-ph/9806419].


\bibitem{Shih:1998pb}
  H.~H.~Shih, S.~C.~Lee and H.~n.~Li,
  Phys.\ Rev.\  D {\bf 59}, 094014 (1999)
  [arXiv:hep-ph/9810515].


\bibitem{Shih:1999yh}
  H.~H.~Shih, S.~C.~Lee and H.~n.~Li,
  Phys.\ Rev.\  D {\bf 61}, 114002 (2000)
  [arXiv:hep-ph/9906370].


\bibitem{Guo:2005qa}
  P.~Guo, H.~W.~Ke, X.~Q.~Li, C.~D.~Lu and Y.~M.~Wang,
  Phys.\ Rev.\  D {\bf 75}, 054017 (2007)
  [arXiv:hep-ph/0501058].



\bibitem{He:2006ud}
  X.~G.~He, T.~Li, X.~Q.~Li and Y.~M.~Wang,
  Phys.\ Rev.\  D {\bf 74}, 034026 (2006)
  [arXiv:hep-ph/0606025].



\bibitem{Huang:1998rq}
  C.~S.~Huang, C.~F.~Qiao and H.~G.~Yan,
  Phys.\ Lett.\  B {\bf 437}, 403 (1998)
  [arXiv:hep-ph/9805452].



\bibitem{Huang:1998ek}
  C.~S.~Huang and H.~G.~Yan,
  Phys.\ Rev.\  D {\bf 59}, 114022 (1999)
  [Erratum-ibid.\  D {\bf 61}, 039901 (2000)]
  [arXiv:hep-ph/9811303].



\bibitem{Colangelo:2000dp}
  For a recent review, see P.~Colangelo and A.~Khodjamirian,
  arXiv:hep-ph/0010175.


\bibitem{Huang:2004vf}
  M.~Q.~Huang and D.~W.~Wang,
  Phys.\ Rev.\  D {\bf 69}, 094003 (2004)
  [arXiv:hep-ph/0401094].



\bibitem{Wang:2008sm}
  Y.~M.~Wang, Y.~Li and C.~D.~Lu,
  Eur.\ Phys.\ J.\  C {\bf 59}, 861 (2009)
  [arXiv:0804.0648 [hep-ph]].



\bibitem{Khodjamirian:2005ea}
  A.~Khodjamirian, T.~Mannel and N.~Offen,
  Phys.\ Lett.\  B {\bf 620}, 52 (2005)
  [arXiv:hep-ph/0504091].




\bibitem{Khodjamirian:2006st}
  A.~Khodjamirian, T.~Mannel and N.~Offen,
  Phys.\ Rev.\  D {\bf 75}, 054013 (2007)
  [arXiv:hep-ph/0611193].


\bibitem{Faller:2008tr}
  S.~Faller, A.~Khodjamirian, C.~Klein and T.~Mannel,
  Eur.\ Phys.\ J.\  C {\bf 60}, 603 (2009)
  [arXiv:0809.0222 [hep-ph]].



\bibitem{Wang:2008bw}
  Z.~G.~Wang,
  Phys.\ Lett.\  B {\bf 666}, 477 (2008)
  [arXiv:0804.0907 [hep-ph]].



\bibitem{Wang:2009ym}
  Z.~G.~Wang,
  arXiv:0906.4206 [hep-ph].




\bibitem{Ball:2008fw}
  P.~Ball, V.~M.~Braun and E.~Gardi,
  Phys.\ Lett.\  B {\bf 665}, 197 (2008)
  [arXiv:0804.2424 [hep-ph]].



\bibitem{Lange:2003ff}
  B.~O.~Lange and M.~Neubert,
  Phys.\ Rev.\ Lett.\  {\bf 91}, 102001 (2003)
  [arXiv:hep-ph/0303082].



\bibitem{Lepage:1979zb}
  G.~P.~Lepage and S.~J.~Brodsky,
  Phys.\ Lett.\  B {\bf 87}, 359 (1979).


\bibitem{asymptonic forms 1}I.I.\ Balitsky and V.M.\ Braun, Nucl.\
Phys.\ B  {\bf 311} (1989) 541.




\bibitem{Khodjamirian:1998ji}
  A.~Khodjamirian and R.~Ruckl,
  Adv.\ Ser.\ Direct.\ High Energy Phys.\  {\bf 15} (1998) 345
  [arXiv:hep-ph/9801443].



\bibitem{higher Fock state}M.~Diehl, T.~Feldmann, R.~Jakob and P.~Kroll,
  Eur.\ Phys.\ J.\  C {\bf 8} (1999) 409
  [arXiv:hep-ph/9811253].


\bibitem{Amsler:2008zzb}
  C.~Amsler {\it et al.}  [Particle Data Group],
  Phys.\ Lett.\  B {\bf 667}, 1 (2008).



\bibitem{ioffe}B.L. Ioffe, Prog. Part. Nucl. Phys. {\bf{56}}, 232
 (2006)  [arXiv: hep-ph/0502148].


\bibitem{Braun:2006hz}
  V.~M.~Braun, A.~Lenz and M.~Wittmann,
  Phys.\ Rev.\  D {\bf 73}, 094019 (2006)
  [arXiv:hep-ph/0604050].


  \bibitem{Liu:2008yg}
  Y.~L.~Liu and M.~Q.~Huang,
  Nucl.\ Phys.\  A {\bf 821}, 80 (2009)
  [arXiv:0811.1812 [hep-ph]].



\bibitem{dosch}
H.G. Dosch, E.M. Ferreira,  F.S. Navarra and  M. Nielsen,  Phys.
Rev. {\bf{D 65}} (2002)  114002 [arXiv:hep-ph/0203225].

\bibitem{matheus}R.D. Matheus, F.S. Navarra, M. Nielsen and R. Rodrigues da
Silva,  Phys. Lett. {\bf{B 541}} (2002) 265  [arXiv:hep-ph/0206198].

\bibitem{bracco}M.E. Bracco, M. Chiapparini, F.S. Navarra and M.
Nielsen, Phys. Lett. {\bf{B 605}} (2005) 326 [arXiv:hep-ph/0410071].


\bibitem{navarra}F.S. Navarra, Marina Nielsen, M.E. Bracco,
M. Chiapparini and C.L. Schat, Phys. Lett. {\bf{B 489}} (2000) 319
[arXiv:hep-ph/0005026].



\bibitem{Wang:2007ys}
  Y.~M.~Wang, H.~Zou, Z.~T.~Wei, X.~Q.~Li and C.~D.~Lu,
  Eur.\ Phys.\ J.\  C {\bf 54} (2008) 107
  [arXiv:0707.1138 [hep-ph]].



\bibitem{Colangelo}P. Colangelo, G. Nardulli and N. Paver, Z. Phys. {\bf{C 57}},
43 (1993).








\bibitem{Braun:2001tj}
  V.~M.~Braun, A.~Lenz, N.~Mahnke and E.~Stein,
  Phys.\ Rev.\  D {\bf 65}, 074011 (2002)
  [arXiv:hep-ph/0112085].









\bibitem{Lu:2009cm}
  C.~D.~Lu, Y.~M.~Wang, H.~Zou, A.~Ali and G.~Kramer,
  arXiv:0906.1479 [hep-ph].






\end{thebibliography}
\end{document}